# Ultra-fast Oxygen Conduction in Sillén Oxychlorides

Jun Meng*, Md Sariful Sheikh, Lane E. Schultz, William O. Nachlas, Jian Liu, Maciej P. Polak, Ryan Jacobs, Dane Morgan*

J. Meng, M. S. Sheikh, L. E. Schultz, M. P. Polak, R. Jacobs, D. Morgan. *Department of Materials Science and Engineering, University of Wisconsin-Madison, 1509 University Ave. Madison, Wisconsin 53706, United State. Emails :* jmeng43@wisc.edu; ddmorgan@wisc.edu

W. O. Nachlas. *Department of Geoscience, University of Wisconsin-Madison, 1215 West Dayton Street, Madison, Wisconsin 53706, United State.*

J. Liu. *National Energy Technology Laboratory, 3610 Collins Ferry Road, Morgantown, West Virginia 26505, United State.*

**Data Availability Statement:** The data that support the findings of this study are available on Figshare: https://figshare.com/s/4a54f0be848757ae8aee.

**Funding Statement:** This work was funded by the US Department of Energy (DOE), Office of Science, Basic Energy Sciences (BES), under Award # DE-SC0020419. This work used the computational resources from the Advanced Cyberinfrastructure Coordination Ecosystem: Services & Support (ACCESS) program, which is supported by National Science Foundation grants #2138259, #2138286, #2138307, #2137603, and #2138296.

**Conflict of Interest:** The authors declare no conflict of interest.

# Abstract

Oxygen ion conductors are crucial for enhancing the efficiency of various clean energy technologies, including fuel cells, batteries, electrolyzers, membranes, sensors, and more. In this study, LaBi$_2$O$_4$Cl is identified as an ultra-fast oxygen conductor from the MBi$_2$O$_4$X (M=rare-earth element, X=halogen element) family, discovered by a structure-similarity analysis of >60k oxygen-containing compounds. *Ab initio* studies reveal that LaBi$_2$O$_4$Cl has an ultra-low migration barrier of 0.1 eV for oxygen vacancy, significantly lower than 0.6-0.8 eV for interstitial oxygen. Frenkel pairs are the dominant defects in intrinsic LaBi$_2$O$_4$Cl, facilitating notable oxygen diffusion primarily through vacancies at higher temperatures. LaBi$_2$O$_4$Cl with extrinsic oxygen vacancies (2.8%) exhibits a conductivity of 0.3 S/cm at 25°C, maintains a 0.1 eV diffusion barrier up to 1100°C, and transitions from extrinsic to mixed extrinsic and intrinsic behavior as the Frenkel pair concentration increases at higher temperatures. Experimental results on synthesized LaBi$_2$O$_4$Cl and Sr-doped LaBi$_2$O$_4$Cl demonstrate comparable or higher oxygen conductivity than YSZ and LSGM below 400 °C, with lower activation energies. Further experimental optimization of LaBi$_2$O$_4$Cl, including aliovalent doping and microstructure refinement, could significantly enhance its performance and efficiency, facilitating fast oxygen conduction approaching room temperature.

# Introduction

Solid state ionic conductors play a crucial role in numerous energy device technologies, offering a range of benefits and applications, such as fuel cells,[1–4] traditional[5,6] and new solid-state batteries,[7–10] electrolyzers,[11] sensors,[12,13] membranes,[14] memristors,[15–17] and more. By enabling the movement of ions in robust structures, solid ionic conductors allow for the effective conversion of chemical energy directly into electrical energy and vice versa. The discovery of new ionic conductors has the potential to improve the performance and viability of many energy devices, enabling faster transport and reaction rates, higher power densities, improved device efficiency, and, through these improved properties, more attractive economic incentives for their more widespread use and adoption.

Oxygen is one of the most useful ions to be able to conduct, and discovering new fast solid state oxygen ion conducting materials is particularly important for improving efficiency of fuel cells, including traditional and reversible solid oxide fuel cells (SOFCs, r-SOFCs) and proton ceramic fuel cells (PCFCs), solid oxide electrolysis cells (SOECs), solid oxide air batteries, oxygen gas sensors, and oxygen separation membranes. Over the past several decades, significant progress has been made in identifying and characterizing solid state oxygen conductors, which can be classified into several groups based on their structural characteristics. These groups include (1) fluorite-related oxides like yttria-stabilized zirconia (YSZ), gadolinia-doped ceria (CGO),[18] Bi$_2$O$_3$,[19] and pyrochlore-type oxides,[20] (2) perovskite oxides such as La$_{0.9}$Sr$_{0.1}$Ga$_{0.8}$Mg$_{0.2}$O$_{3-\delta}$ (LSGM),[21,22] and perovskite structure derivatives like layered perovskite,[23] Brownmillerite,[24] Ruddlesden-Popper,[25] and hexagonal perovskite,[26,27] (3) Aurivillius-type materials,[28–30] consisting of fluorite-like and perovskite-like layers, (4) apatite-type oxides,[31,32] (5) mellite-type oxides,[33,34] (6) scheelite-type oxides,[35,36] (7) mayenite,[37] (8) hexagonal manganites,[38] and (9) perrierite-type oxides.[39] At present, the known oxygen conductors have demonstrated substantial oxygen ionic conductivity ≈ 0.1 S/cm at intermediate temperatures around 600 °C or higher. Typical oxygen-conducting perovskite materials have an activation barrier of around 1 eV, which incurs a reduction in conductivity upon cooling of about 10x per 100 K. Modern electrochemical devices approximately target about 1 A/cm$^2$ currents, 10$^{-2}$ cm materials thickness, and < 0.1 V Ohmic losses from ionic transport, which requires an ionic conductivity of 0.1 S/cm or greater. Therefore, present oxygen conductivities suggest that it is difficult to utilize oxygen conducting materials in devices operating much below 600 °C. This high operating temperature can result in high system costs, materials degradation, and sluggish start-up and shutdown cycles. Finding new materials with high oxygen ionic conductivity at significantly lower temperatures (e.g., room temperature to 400 °C) and reduced activation energies for oxygen transport could significantly improve oxygen-based electrochemical technologies.

The relatively recent growth of large, curated materials databases and the maturation of data-centric techniques now enables unprecedented search and screening of new materials. Recently, Zhang et al. used unsupervised machine learning methods utilizing the structure of the anion framework of materials to discover numerous new structure types of solid-state Li ion conductors.[40] Inspired by this approach, we analyze ≈ 62k O-containing materials in the Materials Project[41] to search for new fast oxygen conducting materials. Our approach is based on the fundamental hypothesis that the structure of a material is the key factor in governing its oxygen ion conduction characteristics. In this work, we featurize the structure of each O-containing material using the simulated X-ray diffraction (XRD) pattern of the O-sublattice and radial distribution function (RDF). Then, we perform a structure similarity analysis by calculating the Euclidean distance of these XRD- and RDF-derived features of each O-containing structure relative to a reference structure. Here, fluorite $BiO_2$ (Materials Project entry mp-32548) was chosen as the reference structure due to its prevalence in excellent oxygen conductors such as YSZ and CGO. We use the structural similarities to fluorite to generate a prioritized list of materials, excluding those that are structurally equivalent to the well-known fluorite-type oxygen conductors such as $ZrO_2$ and $Bi_2O_3$. Using this structure similarity analysis, together with subsequent screening of material stability and synthesizability, we identified the $Bi_2MO_4X$ (M=rare-earth element, X=halogen element) structure family as a new group of fast oxygen conductors. Additional details of this analysis are given in the **Methods 1** and **Supplementary Information (SI) Section 1**.

Among the $MBi_2O_4X$ family, $LaBi_2O_4Cl$ (LBC) was selected as a representative composition for in-depth studies due to its straightforward synthesis procedure and the low price of La compared to other rare-earth elements. LBC was first reported as a novel variant of the Sillén phase in 1995.[42] Nakada *et al.* recently reported that LBC has a bandgap of 2.79 eV and presents promising visible-light photocatalytic activity for water splitting.[43] However, the oxygen mobility in this structure had not been studied until very recently. While this work was in progress,[44] a concurrent study conducted by Yaguchi *et al.* reported fast oxygen ion conduction in Te-doped LBC, attributed to an interstitial-mediated diffusion mechanism.[45] Interestingly, our *ab initio* studies have identified that LBC has the potential to be an ultra-fast oxygen conductor with oxygen transport occurring via a vacancy-mediated diffusion mechanism, with an ultra-low migration barrier of 0.1 eV, while we predict interstitial conduction mechanisms to have higher activation energies in the range of 0.6-0.8 eV. Here, both *ab initio* studies and experiments were conducted to study the crystal structure, electrical properties, defect chemistry, and oxygen ion mobility in LBC. From *ab initio* studies, we discovered that O-vacancy containing LBC demonstrates a notably low vacancy-mediated oxygen ion diffusion barrier of 0.1 eV below 1100 °C and has extremely high predicted ionic conductivity of 0.3 S/cm at 25 °C with 2.8% oxygen vacancies (concentration give as fraction of O sites). We also identified that the Frenkel pair is the most stable intrinsic defect, contributing to fast oxygen ion conduction in intrinsic LBC at high temperatures, with a conductivity of 0.1 S/cm at 950 °C at a Frenkel pair concentration of approximately 0.34 % (given as a percentage of oxygen involved in Frenkel pair formation), which is predicted to occur spontaneously in intrinsic LBC due to thermal excitations. Our experimental conductivity shows LBC and Sr-doped LBC present comparable conductivity to other state-of-art oxygen conductors like YSZ, CGO and LSGM, but with an experimental conductivity more similar to our simulated intrinsic LBC material than the very high conductivity values predicted for O vacancy-containing LBC. Our findings suggest that the true O conduction potential of LBC has not yet been realized, and we propose that the conductivity of LBC at low temperature can be enhanced by creation of extrinsic vacancies (e.g., through aliovalent doping) and through microstructural engineering (e.g., through grain refinement).

## Results and Discussion
### Structure, electronic properties and oxygen defect chemistry in LaBi$_2$O$_4$Cl

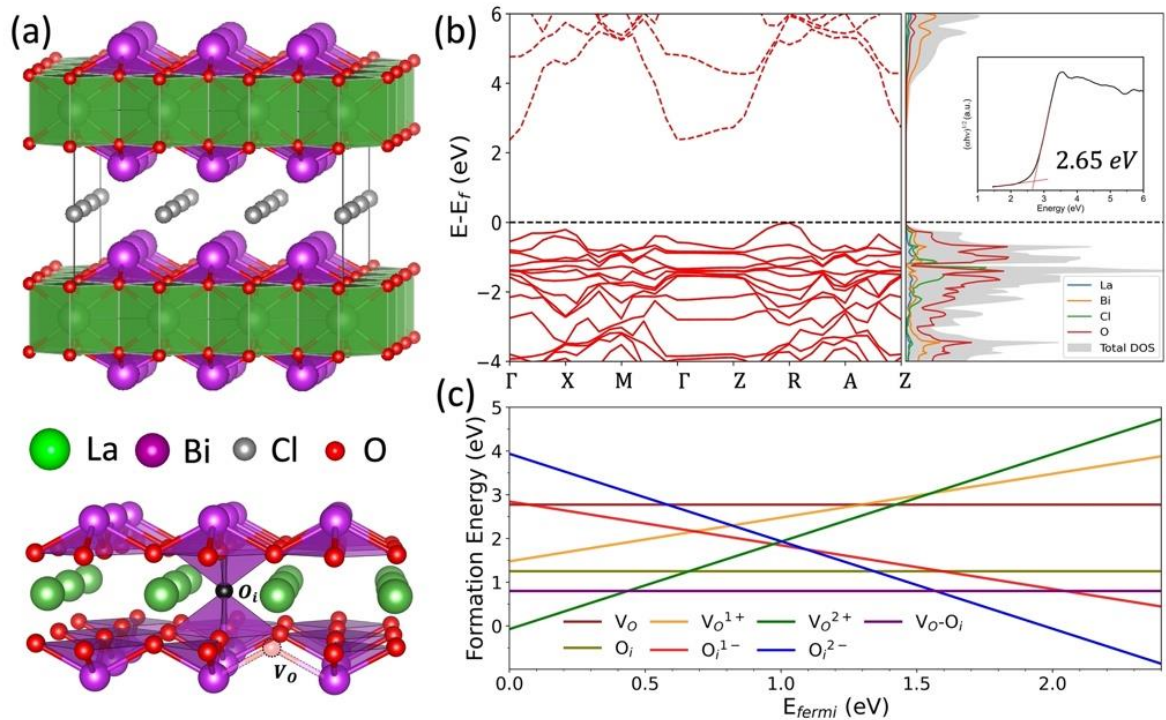

**Figure 1.** (a) Structure of LaBi$_2$O$_4$Cl (LBC). The La, Bi, Cl and O atoms are shown as green, purple, grey, and red spheres, respectively. (b) Band structure and density of states of bulk LBC calculated with the HSE06 functional. Inset picture shows the optical bandgap determination from the UV-vis diffuse reflectance spectroscopy of the in-house synthesized LBC pellet. (c) Defect formation energies as a function of Fermi level for intrinsic oxygen defects considered in LBC at 300K under atmospheric $P_{O_2}$. The defect formation energies were calculated with generalized gradient approximation exchange-correlation functional Perdew, Burke, and Ernzerhof (GGA-PBE),[46] and the modified band alignment (MBA) method[47] was used to shift the bandgap and defect states to better represent the true physics of the material (see **Methods 2**).

LBC is a layered bismuth-based oxyhalide, which crystallizes in the space group P4/mmm, and has some similar structural characteristics to fluorite but is still unique from any known fast oxygen conductor. As shown in **Figure 1a**, LBC features a "triple fluorite" layer [LaBi$_2$O$_4$]$^+$, which is charge-balanced by the anion Cl$^-$ layers. Within the "triple fluorite" layer, the edge-sharing LaO$_2$ cubes are sandwiched by Bi$_2$O$_2$ square pyramids on both sides. This 2D layered structure offers well-defined anisotropic ionic diffusion pathways in which ions can move in the triple fluorite layers. Notably, many materials with similar 2D layered structural frameworks have demonstrated rapid ionic conduction, such as O diffusion in Ruddlesden-Popper phases La$_{2-x}$Sr$_x$NiO$_4$[25,48] and Li conduction in layered LiCoO$_2$ and related phases.[49] The *ab initio* calculated bandgap is 2.38 eV using the hybrid functional Heyd, Scuseria, and Ernzerhof (HSE06),[50] consistent with the experimental value of 2.65 eV from UV-vis diffuse reflectance measurements on synthesized LBC shown in **Figure 1b** and **Table S2**.

The thermodynamics of intrinsic oxygen point defects in LBC was investigated by *ab initio* studies. The focus was on three primary defects: oxygen vacancy ($V_O$), oxygen interstitial ($O_i$), and Frenkel pair ($V_O - O_i$), and these defect sites are displayed in **Figure 1a**. In the following discussion, the charge states of point defects are given with relative charge using Kröger–Vink notation, where the net charge on defects was added to the unit cell in the *ab initio* calculations (see **SI Sec. 2.1**). Charge states of 0 ($V_O$), 1+ ($V_O^\cdot$), and 2+ ($V_O^{\cdot\cdot}$) were investigated for oxygen vacancies, and charge states of 0 ($O_i$), 1- ($O_i'$), and 2- ($O_i''$) were investigated for oxygen interstitials. Defect formation energies were calculated with the generalized gradient approximation exchange-correlation functional Perdew, Burke, and Ernzerhof (GGA-PBE),[46] which underestimates the bandgap value. To circumvent this issue, we adopted a recently developed modified band alignment (MBA) method[47] to shift the band edges and defect levels from the GGA-PBE ($E_{gap} = 1.28\ eV$) level to more closely (though not exactly) reflect those present at the HSE level ($E_{gap} = 2.38\ eV$). This approach is useful to have semi-quantitative defect charge state transition levels without having to conduct computationally expensive HSE calculations. **Figure 1c** shows the defect formation energy versus the Fermi energy ($E_{fermi}$) of LBC after band alignment (values are given in **Table S3**). Considering the fundamental principle that charge neutrality of intrinsic defects must be preserved, the equilibrium position of $E_{fermi}$ for the intrinsic material should reside near the crossover point where the lowest energy defect types of opposite charge states that have the same formation energy. In the case of LBC, the defect formation energy of $V_O^{\cdot\cdot}$ and $O_i''$ meet at $E_{fermi} = 1.0\ eV$ above the valence band maximum. At this equilibrium position, the Frenkel pair is predicted to be the most stable intrinsic defect type due to it having the lowest formation energy of the defect types considered. When only considering intrinsic defects, the Frenkel pair persists as the dominant defect type across an extensive temperature range from room temperature to temperatures up to at least 800°C (**Figure S1**). We believe the preference for forming Frenkel pair defects in LBC, rather than vacancies or interstitials, is due to the material's redox properties and sizable bandgap. In LBC, reducing La$^{3+}$ and/or Bi$^{3+}$ to lower valence states (e.g., Bi$^{2+}$) is energetically unfavorable, making oxygen vacancy creation (electron doping) challenging. This observation is supported by our density of states analysis shown in **Figure S2**, where creating the $V_O$ and $V_O^\cdot$ defects result in a significant upward shift of the Fermi level towards the conduction band. This shift of Fermi level, caused by doping electrons at high energy levels near the conduction band, is the energy cost which increases the overall formation energy of these defects. Although $Bi^{3+}$ can be oxidized to higher valence states such as $Bi^{4+}$ or $Bi^{5+}$, these states are relatively unstable. Thus, even though generating an oxygen interstitial (hole doping) requires less energy than creating an oxygen vacancy, it still costs 1.25 eV for forming an oxygen interstitial at room temperature atmosphere (**Table S3**), making it energetically costly. Consequently, the formation of Frenkel pairs remains the most energetically favorable defect. We show in the next section that the formation of Frenkel pairs occurs spontaneously at high temperature and leads to notable O conduction at high temperature in *ab initio* simulation of intrinsic LBC.

Oxygen mobility in LaBi$_2$O$_4$Cl

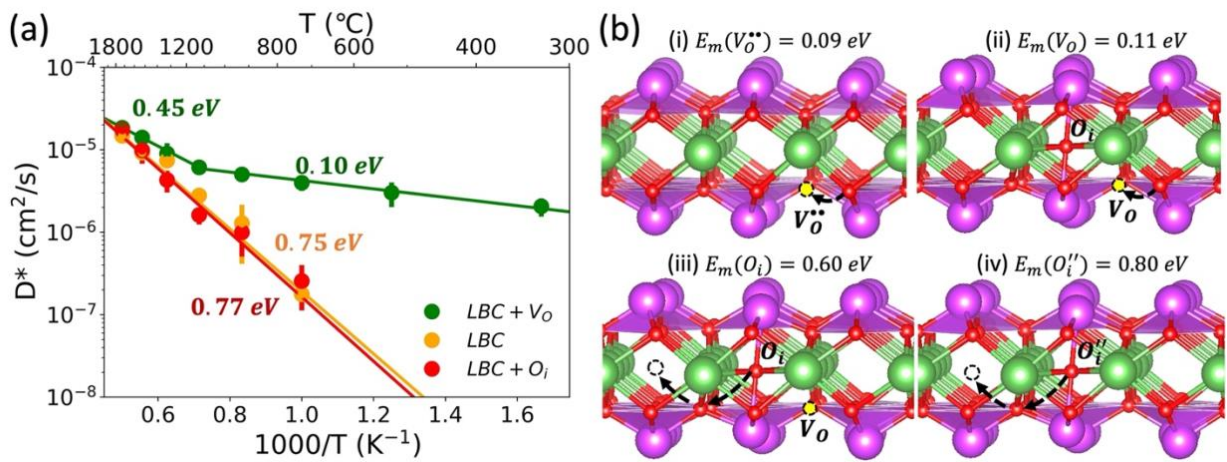

**Figure 2.** (a) Arrhenius plot of oxygen ion self-diffusivity $D^*$ in LBC, LBC with 2.8% extrinsic oxygen vacancies, and LBC with 2.8% extrinsic oxygen interstitials, calculated from *ab initio* molecular dynamic simulations. (b) Illustrations of the vacancy-mediated diffusion mechanism in the presence of (i) just oxygen vacancy $V_O^{\bullet\bullet}$, (ii) a Frenkel pair ($V_O - O_i$), and interstitial-mediated diffusion mechanism in the presence of (iii) a Frenkel pair, (iv) interstitial $O_i''$. The migration barriers are calculated by the climbing image nudged elastic band (CI-NEB) method. The diffusion pathways are represented by the dashed arrows.

Oxygen ion mobility in the presence of various oxygen defects was examined through *ab initio* molecular dynamics (AIMD) simulations. We assessed the self-diffusivity $D^*$ of oxygen by calculating mean squared displacement (MSD) (details in **Methods 2**). **Figure 2a** presents the self-diffusivity $D^*$ of oxygen species at different temperatures for LBC with extrinsic oxygen vacancies (LBC+$V_O^{\bullet\bullet}$), intrinsic LBC, and LBC with extrinsic interstitial oxygen (LBC+$O_i''$), respectively. A fractional concentration of approximately 2.8% (or 1/36, which is one defect in a $3 \times 3 \times 1$ supercell) for either vacancies or interstitials was examined. The diffusion mechanisms identified in AIMD simulation were further analyzed using Climbing Image Nudged Elastic Band (CI-NEB) calculations to assess their migration barriers, as shown in **Figure 2b**. The vacancy-mediated diffusion mechanism involves oxygen ions hopping through vacancies in the Bi$_2$O$_2$ layer with a hop distance of 2.8 Å. The migration barriers were determined to be 0.09 eV for extrinsic defect $V_O^{\bullet\bullet}$ and 0.11 eV for $V_O$ associated with a Frenkel pair. The interstitial-mediated (i.e., interstitialcy) diffusion mechanism involves an interstitial oxygen displacing a lattice oxygen to a new interstitial site. The migration barrier for this interstitial-mediated mechanism is found as 0.60 eV for $O_i$ associated with a Frenkel pair and 0.80 eV for extrinsic defect $O_i''$.

For intrinsic LBC (yellow data in **Figure 2a**), the spontaneous formation of Frenkel pairs is observed in AIMD studies at 1000K and above within a total simulation time of about 400-800 ps. The oxygen diffusion is dominated by a vacancy-mediated mechanism and the overall activation barrier is 0.75 ± 0.01 eV, derived by fitting the Arrhenius relation. Using the Nernst-Einstein relation (**Methods 2**), our simulations predict an ionic conductivity of 0.1 S/cm around 950°C for intrinsic LBC, within the range of current best fast oxygen conductors discussed in the following section.

For LBC with interstitial oxygen (LBC+$O_i''$) (red data in **Figure 2a**), the energy barrier of $O_i''$ diffusion via an interstitial-mediated pathway (**Figure 2b** (iv)) is calculated to be 0.80 eV using CI-NEB. This energy barrier is comparable to the formation energy of a Frenkel pair, which is calculated to be 0.79 eV at T= 0 K, indicating that $O_i''$ diffusion and Frenkel pair formation will occur approximately together. However, oxygen diffusion will primarily occur through oxygen vacancies once a Frenkel pair forms, as oxygen vacancies have a significantly lower migration barrier than interstitials. AIMD studies further confirm that Frenkel pair formation in LBC+$O_i''$ and oxygen ion diffusion through vacancy-mediated mechanism at 1000K and above occur within a total simulation time of about 460-780 ps. Our studies show that oxygen diffusion in LBC+$O_i''$ system is dominated by vacancy-mediated diffusion from Frenkel pairs, despite the presence of additional $O_i''$. This mechanism results in a total activation energy of 0.77 ± 0.06 eV, which closely aligns to intrinsic LBC, as both are dominated by the same defect formation and migration mechanism. In addition, the fact that LBC+$O_i''$ gives similar values of diffusivities and activation energy as intrinsic LBC suggests that excess oxygen interstitials do not significantly impact the oxygen mobility in LBC. However, the extra interstitial does increase the overall conductivity due their significant concentration, and the LBC+$O_i''$(with 2.8% extrinsic interstitials) is predicted to have an ionic conductivity of 0.1 S/cm at 850 °C, about compared to 950 °C noted above for the intrinsic LBC.

For LBC with oxygen vacancy (LBC+$V_O^{..}$) (green data in **Figure 2a**), distinct diffusion behaviors are observed at different temperature regimes, attributed to the presence of different defect types. At low temperatures, $V_O^{..}$ is the dominant defect, and oxygen ions diffuse through the vacancy-mediated mechanism. This low temperature diffusion regime is characterized by an ultra-low migration barrier of 0.10 ± 0.02 eV, in excellent agreement with the CI-NEB-derived barrier of 0.09 eV (**Figure 2b** (i)). As temperature increases, Frenkel pairs begin to form, creating new fast diffusing vacancies as well as some new but slower diffusing interstitials. The additional defects created by spontaneous Frenkel pair formation begin to impact the overall diffusion at about 1100 °C, changing the slope of self-diffusivity and leading to an increased activation energy of 0.45 ± 0.07 eV, as shown in **Figure 2a**. With the presence of about 2.8% extrinsic oxygen vacancies, LBC is predicted to achieve conductivity of 0.3 S/cm at 25 °C.

The above qualitative analysis of the transition of active defects and mechanisms in LBC+$V_O^{..}$ is made more quantitative in **SI Section 2.3**. There we develop a simple model to calculate the overall diffusivity versus temperature, assuming non-interacting diffusing species and an Arrhenius form for Frenkel pair concentration. The model uses the following attempt frequencies and migration barriers for the defects: $f_v$=7×10$^{11}$ s$^{-1}$, $f_i$=5×10$^{12}$ s$^{-1}$ and $E_m^v =$ $0.10\ eV$, $E_m^i = 0.60\ eV$ for vacancies (*v*) and interstitials (*i*), respectively. The model also uses a fitted Frenkel pair formation energy of 0.60 eV, which is 0.19 eV smaller than the value calculated directly from *ab initio* methods at T = 0 K (**Table S3**). This difference is likely due to approximations in the simple model (e.g., excluding Frenkel pair formation entropy) and temperature-induced structural distortions that effectively lower the Frenkel pair formation energy in AIMD simulation. This simple model predicts diffusivities and activation energies that quantitatively match the AIMD results across all the three cases (LBC+$V_O^{\cdot\cdot}$, intrinsic LBC, and LBC+$O_i''$, as shown in **Figure S3**). Thus this model quantitatively shows how changes in active defects and mechanisms affect the overall slope of the diffusivity versus temperature.

The ultra-low migration barrier for oxygen vacancy diffusion suggests LBC could exhibit fast oxygen ion conduction near room temperature. Such performance would likely be realized by doping the LBC to create vacancies, but dopant-vacancy binding or other effects that reduce oxygen ion mobility are a concern in such an approach. To assess if doping might help realize the potential of this system, we first studied the effects of p-type aliovalent doping to stabilize oxygen vacancies in LBC through *ab initio* studies. We chose $Sr^{2+}$ as a replacement for $La^{3+}$ to create extrinsic oxygen vacancies, resulting in Sr-doped LBC $La_{0.78}Sr_{0.22}Bi_2O_{3.89}Cl$ (SLBC), where the formation of $V_O$ becomes thermodynamically favorable under atmosphere (**Table S3**). AIMD studies revealed that $LaBi_2O_{3.89}Cl$ and $La_{0.78}Sr_{0.22}Bi_2O_{3.89}Cl$ exhibit similar oxygen self-diffusivity values and trends in **Figure S4**. Two distinct activation energies are observed for $La_{0.78}Sr_{0.22}Bi_2O_{3.89}Cl$ at different temperature regimes, mirroring the behavior of the Sr-free material. Below 1100 °C, oxygen ion diffusion primarily occurs through a vacancy-mediated mechanism, with a migration barrier of 0.16 ± 0.04 eV, consistent with CI-NEB calculated values (0.11-0.27 eV) for all considered diffusion pathways (**Figure S6**). The binding energy between oxygen vacancies and Sr dopants is relatively small (0.08, 0.11, and 0.19 eV in **Figure S6**), consistent with relatively weak trapping of oxygen vacancies by the Sr dopants. As the temperature rises, Frenkel pairs once again form, leading to an increase in defect concentration and subsequently enhancing oxygen self-diffusivity, albeit with a higher activation energy of 0.53 ± 0.08 eV. The simulated oxygen-deficient Sr-doped LBC achieves conductivity higher than 0.1 S/cm over a broad temperature range from 800°C down to 100°C (discussed below), demonstrating potentially outstanding low temperature oxygen transport.

## Experimental evaluation of oxygen conduction in LBC

To experimentally investigate the oxygen conductivity in LBC and Sr-doped LBC, we synthesized pure LBC powder and Sr-doped LBC powder using a flux synthesis method.[43] X-ray diffraction (XRD) analysis indicated that the as-synthesized LBC powder exhibits a monoclinic phase, which then transformed to a tetragonal phase after sintering in pellet form at 950°C for 10 hours in flowing Ar gas (**Figure 3a**). The obtained lattice parameters of LBC were found to be consistent with the *ab initio* calculation results and previous reports[43,51] of monoclinic and tetragonal LBC (**Table S4**). **Figure 2Figure 3b** displays the field emission scanning electron microscopy (FESEM) image of the LBC pellet surface, showing a dense surface and an average grain size of 2.8 μm. The stoichiometry of each element in LBC was evaluated by Electron probe microanalysis (EPMA) studies, suggesting a chemical formula of $La_{0.96}Bi_2O_{3.91}Cl_{1.05}$, with slight La and oxygen deficiencies, and excess Cl, in the synthesized LBC pellet (**Table S5**). Sr-doped LBC (SLBC) was synthesized with the aim of including 5% Sr dopants and generating oxygen vacancies. The XRD analysis on SLBC shows that the as-synthesized powder also adopted the monoclinic phase, and the sintered pellet adopted the tetragonal phase (**Figure S7**). The results indicate that SLBC presents a slight lattice expansion (less than 1%) compared to LBC, consistent with expectations from Sr doping, as the ionic radius of $Sr^{2+}$ is larger than $La^{3+}$. EPMA studies show that SLBC has a chemical formula of $La_{0.89}Sr_{0.01}Bi_2O_{3.89}Cl_{1.02}$, indicating only about 1% Sr incorporated into LBC. It should be noted that the chemical formula exhibits the average stoichiometry of each element, which may not always result in a charge-neutral formula. This is attributed to the significant error range in the stoichiometry of each element as measured by EPMA, as shown in **Table S5**. The impact of this error range is discussed below.

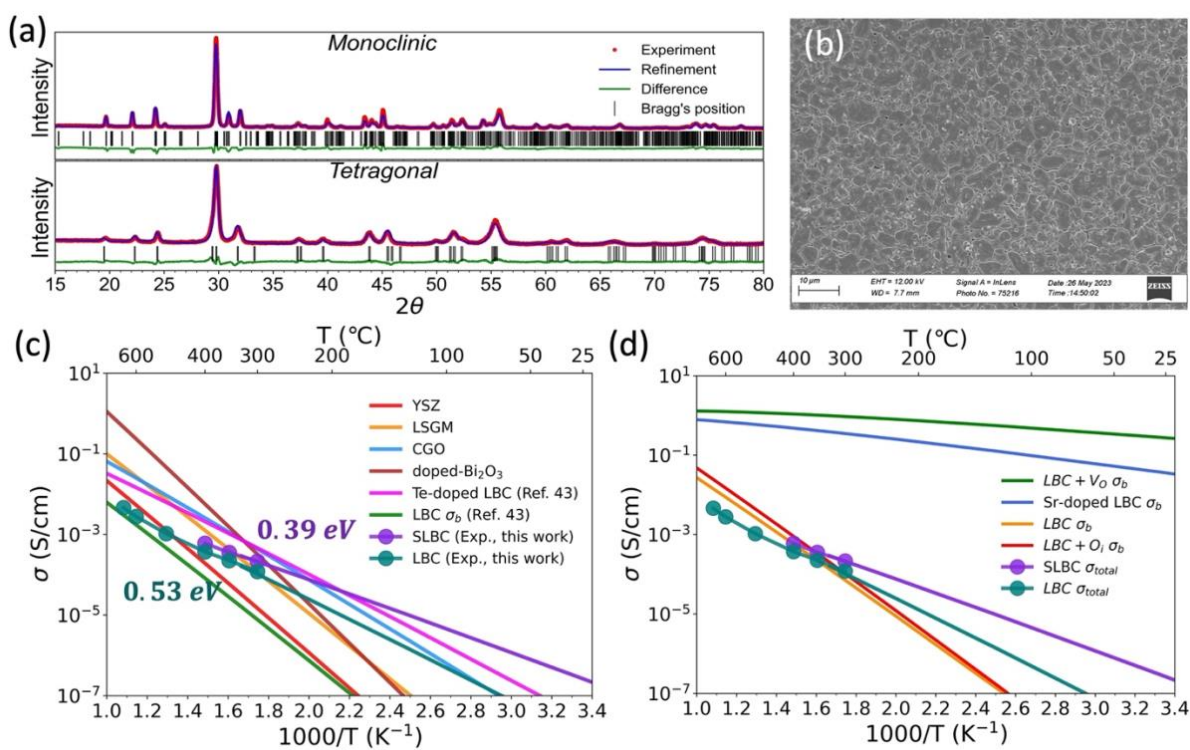

**Figure 3.** (a) X-ray powder diffraction data of the synthesized $LaBi_2O_4Cl$ powder fitted to the monoclinic phase, and the sintered $LaBi_2O_4Cl$ pellet fitted to the tetragonal phase. (b) FESEM image of the $LaBi_2O_4Cl$ pellet surface. (c) The predicted ionic conductivity and experimental total conductivity in intrinsic $LaBi_2O_4Cl$, Sr-doped $LaBi_2O_4Cl$ from this work compared with the state-of-art fast oxygen conductors. (d) The predicted ionic conductivity from AIMD simulation compared with experimental conductivity.

The experimental total conductivity was measured by the DC 4-probe method (**Methods 3**). Given that LBC has a sizeable bandgap of 2.65 eV, and the cations $La^{3+}$ and $Bi^{3+}$ do not readily change valence states, it is unlikely that LBC exhibits significant electronic conductivity, either from thermal excitation of electrons to the conduction band or through polaron hopping. Hence, the electronic conductivity was considered negligible, and the total conductivity was assumed to be due to ionic conductivity. We note that this same assumption was made in the recent study on ionic conductivity of Te-doped LBC.[45] In **Figure 3c**, the total conductivity of LBC is comparable to many state-of-the-art fast oxygen conducting electrolyte materials. Below 400°C, LBC exceeds the performance of the widely used YSZ, is comparable to LSGM, and has lower activation energy than both materials. The experimental conductivity activation energy was found to be 0.53 ± 0.01 eV. Sr-doped LBC displayed approximately 2x improvement compared to LBC, with a reduced activation energy of 0.39 ± 0.01 eV. This improvement might be due to the expected increase in the amount of oxygen vacancies with Sr doping, although more experiments would be needed to verify this hypothesis. The small activation barrier suggests that SLBC is expected to exhibit higher conductivity than the other state-of-the-art materials below about 200 °C, as shown in **Figure 3c**. We also compared the conductivity of our LBC and SLBC samples with that of LBC and Te-doped LBC from the recent study from Yaguchi *et al.*[45] The comparison, shown in **Figure 3c**, reveals that our LBC sample exhibits higher conductivity and a lower activation energy than the previously studied LBC. Although the highest-performing Te-doped LBC sample from the literature exhibits higher conductivity than both LBC and SLBC at temperatures above 200 °C, our LBC and SLBC samples have lower migration barriers than the 0.68 eV activation energy barrier reported for the Te-doped LBC sample. In addition, the experimental activation energy for Te-doped LBC aligns well with our *ab initio* studies predicted activation energy of 0.77 eV for LBC with interstitials, which is consistent with the reported oxygen interstitial conduction in Te-doped LBC.[45] This comparison further suggests that the O-vacancy containing LBC holds high potential of exhibiting fast oxygen conduction near room temperature.

Discussion

**Figure 3d** displays an overall comparison across experimental conductivity to the AIMD predicted conductivity across all studied systems. The experimental LBC presents similar conductivity values and activation energy to the *ab initio* intrinsic LBC results, suggesting the experimental sample is behaving more like intrinsic LBC as opposed to the LBC with extrinsic O vacancies, which was suggested by the EPMA results. Therefore, these results suggest that either (i) negligible extrinsic oxygen vacancies exist in our experimental LBC sample or (ii) oxygen vacancies exist but the full potential of their conductivity is not being observed. For (i), it is possible that fewer oxygen vacancies exist than we expect from the EPMA measurement, which is not unlikely given the significant error range on the stoichiometry of each element. For (ii), if we assume that the EPMA measurements are qualitatively accurate and the lowest oxygen vacancy content is about 1.75%, this still suggests there should be significant oxygen conduction facilitated by extrinsically created oxygen vacancies. Therefore, there is reason to believe that oxygen vacancies exist in the system and their conductivity is much lower than predicted from *ab initio* studies. The discrepancy could be attributed to multiple factors. The *ab initio*-predicted conductivity is the bulk conductivity, which was derived from simulations on the bulk crystal structure without influence from grain boundaries, surfaces, secondary phases or other defects (e.g., other point defects or defect clusters). Here, we consider what we believe to be the most likely origin for the difference in oxygen conductivity from the simulations and experiments, which are effects from grain boundaries. In the recent study from Yaguchi *et al.*, the decomposition of bulk and grain-boundary conductivity in LBC and Te-doped LBC samples suggests that the bulk conductivity can be approximately 3x-350x higher than the total conductivity.[45] In our study, the *ab initio*-predicted bulk conductivity for oxygen-deficient LBC is approximately 270x higher than experimental total conductivity at 650 °C. This difference falls within the range of differences observed for Te-doped LBC due to grain boundary effects. Other factors could be related to strong defect interactions or uncertainty in the *ab initio* calculated migration barriers. As observed in AIMD studies, the activation energy of oxygen ion diffusion is sensitive to the oxygen stoichiometry. A small uncertainty (±0.1 eV) in the activation energy can lead to a significant variation in conductivity, approximately three to four orders of magnitude, near room temperature. Considering these potential impacts between experiments and *ab initio* studies, we conclude that the experimental and *ab initio* results are consistent, despite the fact we have not yet achieved the AIMD-predicted high performance in experiments. Based on the above considerations we propose that two potential avenues to realize the full potential of LBC-based materials for fast oxygen conduction are composition refinement (e.g., aliovalent doping) to create oxygen vacancies and engineering the microstructure to reduce grain boundary effects.

## Conclusion

In this work, we identified a promising low temperature fast oxygen conductor $LaBi_2O_4Cl$ (LBC), belonging to the $MBi_2O_4X$ (M= rare earth, X= halogen) family, which is discovered through a structure-similarity based material discovery analysis on the large database of catalogued oxides in Materials Project. In-depth *ab initio* simulations show that LBC has an ultra-low migration barrier of 0.1 eV for oxygen vacancy-mediated diffusion, in contrast to interstitial-mediated diffusion of 0.6-0.8 eV. The *ab initio* studies identified that Frenkel pairs are the dominant defects in intrinsic LBC and enable notable oxygen diffusion primarily through vacancy-mediated diffusion at higher temperatures. *Ab initio* molecular dynamics studies show that LBC and LBC with oxygen interstitials exhibit similar oxygen diffusivities and activation energies, as the ion diffusion in both materials is dominated by oxygen vacancies originating from Frenkel pair formation. However, extra interstitial does increase the overall conductivity due their significant concentration. In contrast, LBC with oxygen vacancies exhibits two distinct diffusion behaviors below and above 1100 °C. At low temperatures, it features an ultra-low diffusion barrier of 0.10 eV, enabling a conductivity of 0.3 S/cm at 25°C. Experimental studies on synthesized LBC and Sr-doped LBC demonstrate that both LBC and Sr-doped LBC have comparable or higher oxygen conductivity than leading oxygen conducting electrolyte materials YSZ and LSGM at low temperatures (< 400°C), and lower activation energies than these leading materials.

Overall, our findings suggest that the oxygen-deficient LBC can outperform well-established oxygen conductors at target lower temperature operation regimes (400°C to room temperature). We propose two future research directions, aliovalent doping and microstructure refinement for LBC-based materials to fully unlock its potential of fast oxygen conduction approaching room temperature. The successful discovery of $MBi_2O_4X$ material family and identification of LBC as a promising room temperature fast oxygen conductor emphasizes the significant potential for discovering new, efficient conductors and the critical role of lattice structure in ion conduction. This work presents exciting approaches to develop novel oxygen conductors and provides understandings that can advance engineering for LBC to fully leverage its potential for fast oxygen conduction near room temperature.

# Methods

1. Materials discovery of fast oxygen conductors LaBi$_2$O$_4$Cl

To search for new families of fast oxygen conductors, approximately 62,000 oxygen-containing compounds in the Materials Project database were considered. Each compound was featurized for structure similarity analysis using two types of data: the radial distribution function (RDF) and the X-ray diffraction (XRD) pattern of the oxygen sublattice. The structures were queried using the Materials Project API with the pymatgen package.[52] For each structure, the RDF features were calculated through the Python API from OVITO,[53] and the XRD data were acquired through the xrd module in pymatgen.[52] For the XRD property, the 2θ values were evenly gridded from 0.0 to 90.0 inclusive, with 901 data points. As for the RDF property, the pair separation distance was gridded into 100 values ranging from 0.025 to 4.975, inclusive. Taking fluorite BiO$_2$ (Materials Project entry mp-32548) as the reference structure, we use the Euclidean distances between BiO$_2$ and each candidate structure to assess 62k compound distances to BiO$_2$. The Euclidean distance was calculated by $\sqrt{\sum_i (x_i - y_i)^2}$, where $i$ represents the features considered, $x$ and $y$ represent two compared materials. Subsequently, all materials were ranked based on the calculated Euclidean distance, from lowest (most similar to least similar) distance from fluorite BiO$_2$.

Following the prioritization of materials through structure ranking, a set of filters were applied to narrow down the list of materials and screen for the most promising candidates. The screening involved excluding materials not found in the ICSD database and those with an energy above the convex hull ($E_{hull}$) greater than 0 meV/atom, thus reducing false positives and increasing chances of successful synthesis. Additionally, materials were filtered based on electronic properties, discarding those with similar valence electron counts to BiO$_2$ and an oxygen p-band center below -3.0 eV to favor materials conducive to oxygen vacancy formation. These filtering criteria were applied to reduce the list of potential materials and prioritize the selection of new types of potential fast oxygen conducting materials. While these filters were designed to prioritize the discovery of new types of oxygen conducting materials, it is important to acknowledge that this approach is a very coarse screening and may potentially exclude numerous promising materials. The specifics of the selection criteria and filters used are thoroughly explained in the **SI, Section 1.**

2. Density Functional Theory (DFT) Calculations

All Density Functional Theory (DFT) calculations were performed using the Vienna ab Initio Simulation Package (VASP) code.[54] The projector augmented wave method (PAW)[55] and the generalized gradient approximation (GGA) exchange and correlation functional Perdew, Burke, and Ernzerhof (PBE)[46] were used for the effective potential for all atoms. The structure optimization, defect formation energy, and molecular dynamic simulations were performed using GGA-PBE functional. The band structure and density of states were performed with the hybrid functional of Heyd, Scuseria, and Ernzerhof (HSE06),[50] as GGA-PBE often underestimates the bandgap. All calculations were performed with spin polarization enabled. The simulations were performed based on a tetragonal LaBi$_2$O$_4$Cl unit cell, which was fully relaxed and had its volume optimized with the GGA-PBE functional. The optimized unit cell contains 8 atoms, with lattice parameters of $a = b = 3.99$ Å, $c = 9.28$ Å, $\alpha = \beta = \gamma = 90°$. The formation energies were calculated in a 4×4×1 supercell structure containing 128 atoms. One vacancy, interstitial, or Frenkel pair was created, corresponding to a defect concentration of approximately 1.5% (fractional concentration) of $V_O$ and 1.5% of $O_i$, respectively. The stopping criteria for total energy calculations were 0.1 meV/cell for electronic relaxation and 0.05 eV/Å for ionic relaxation, respectively. A plane wave cutoff energy of 520 eV was used in all calculations. The Monkhorst–Pack k-point mesh[56] is 1 × 1 × 2 for the 4×4×1 supercell related to defect formation energy calculations. The GGA-PBE calculated formation energies for charged defects were shifted using a recently developed modified band alignment (MBA) method,[47] which conducts the band alignment dependent on the bandgap value and on the position of defect states within the gap. Detailed computational methods are provided in the **SI, Section 2.1.**

Ab initio molecular dynamics (AIMD) simulations were performed with GGA-PBE functional, using gamma-point-only sampling of k-space. Spin-polarized calculations were performed. AIMD simulations are conducted in a 3x3x1 supercell containing 72 atoms. One $V_O^{\cdot\cdot}$ or $O_i''$ was created, corresponding to a 2.8% (fractional concentration) of vacancy and 2.8% of interstitial, respectively. Charge compensated simulation model was built for $V_O^{\cdot\cdot}$ and $O_i''$ by excluding and including two electrons, respectively. For the case of Sr-doped LBC, two Sr dopants were introduced in the 3×3×1 supercell to create one oxygen vacancy so that the whole system remains charge neutral. Two possible configurations of Sr-dopants were studied, and the most stable one (configuration 1 in **Figure S5**) was chosen for subsequent AIMD simulations. The structure was first heated up to 2000K within 0.3 ps in the NVT ensemble using the Andersen thermostat,[57] and then switched to the Nosé–Hoover thermostat[58,59] for all temperatures for long time simulation. No signs of melting at high temperatures were observed. No effort was made to correct for thermal expansion as it was assumed the effect would be small.

Climbing Image Nudged Elastic Band (CI-NEB) method[60] calculations were performed in a 3x3x1 supercell using GGA-PBE functionals. The plane wave cutoff energy was set as 520 eV. The stopping criteria for total energy calculations were 0.1 meV/cell for electronic relaxation and 0.05 eV/Å for ionic relaxation, respectively. Spin-polarized calculation was performed and Monkhorst–Pack k-point meshes of 2x2x2 was used.

The oxygen diffusivity $D^*$ was evaluated by fitting the Einstein relation $D^* = \frac{1}{2dt}\langle MSD \rangle$, where $\langle MSD \rangle$ is the mean squared displacement of oxygen ions, $d$=2 is the dimension of the system as oxygen diffuses within the "triple fluorite" [LaBi$_2$O$_4$]$^+$ layer, and $t$ is the simulation time. To minimize fitting errors, we adopted a procedure that fits the Einstein relation over a selected time range, which was designed to avoid the ballistic region observed at short simulation times and the issues with poor statistics encountered at longer times.[61] For each temperature, we conducted multiple independent AIMD simulations using different initial seeds. The mean diffusivity and its associated uncertainty from these simulations were calculated to determine the diffusivity for each temperature. More details are in the **SI, Section 2.2.** The migration barrier was obtained by fitting the Arrhenius relationship $D^* = D_0 e^{(-\frac{E_m}{k_bT})}$, where $D_0$ is the pre-exponential factor, $k_b$ is the Boltzmann constant, and $T$ is the temperature. The ionic conductivity was calculated from the diffusion coefficient based on the Nernst-Einstein equation[62] $\sigma = \frac{cz^2F^2}{RT}\frac{D^*}{H_R}$, where $\sigma$ is the conductivity, $z$ is the charge of each oxygen ion (-2), $F$ is the Faraday constant, $R$ is the gas constant, and $T$ is temperature. $c$ is the volume concentration of oxygen species, which are $\frac{35}{V}$, $\frac{36}{V}$, and $\frac{37}{V}$, for LBC with oxygen vacancy, intrinsic

LBC, and LBC with interstitial, respectively. $V$ is the volume of the supercell, and $H_R$ is the Haven ratio. $H_R$ describes the correlation in the motion of charged ions and is calculated by the ratio of tracer diffusion coefficient $D^*$ to the charge diffusion coefficient $H_R = \frac{D^*}{D_\sigma}$. We determined $D_\sigma$ from the mean square displacement of center of mass of all diffusers,[63] and obtained $H_R$ to be 0.40 ~ 0.51 for all systems.

3. Synthesis and Characterization

La$_{1-x}$Sr$_x$Bi$_2$O$_4$Cl (x=0, 0.05) samples were synthesized by the flux synthesis method,[43] with high-purity reagents: Bi$_2$O$_3$ (Alfa Aesar, 99.99 %), BiOCl (ACROS Organics, 97 %), La$_2$O$_3$ (Alfa Aesar, 99.99 %), and SrO (Sigma Aldrich, 99.9 %), and flux CsCl (Thermo Scientific, 99.99 %). The reagents were mixed in stoichiometric ratios and combined at a 5% molar concentration of the target compound in CsCl flux. The mixture was heated in an alumina crucible at 800°C for 20 hours, then washed to remove flux and dried. The synthesized powder was ground in acetone medium for 12 hours using a porcelain mortar and pestle. After drying in the air, the powder was mixed with an aqueous polyvinyl alcohol (PVA) solution with a concentration of 3 weight percent (wt%) and ground for 2 hours. The ground mixture was then pressed into a rectangular pellet under 0.25 GPa using a hydraulic press. The dimension of the pellet was 6 mm x 6 mm with an average thickness of 1.5 mm. The compressed LBC pellet was heated to 600°C at a heating rate of 2°C/min and sintered at 600°C for 10 hours in air in a tube furnace, then the pellet was heated in flowing Ar gas at a flow rate of 200 SCCM in the sealed tube. Finally, the LBC pellet was heated from 600 to 950°C over the course of 5 hours and sintered at 950°C for 10 hours. After sintering, the LBC pellet was cooled to room temperature at a cooling rate of 3°C/min. In case of Sr-doped LBC sample, the whole sintering process was performed in Ar. The sintered pellets were manually polished on all sides using sandpaper (grit size P2500).

The structural characterization of the synthesized powder and sintered pellet was performed using room temperature X-ray diffraction method (Cu-Kα source, Bruker D8 Discovery). Rietveld refinement using the Fullprof code[64] was performed to determine the crystal structure. The field emission scanning electron microscopy (FESEM) image of the pellet surface was collected using a high-resolution microscope (Zeiss 1530). The average grain size is estimated from the size of around 250 individual grains randomly selected from the FESEM image of the pellet surface. The grain size was measured using ImageJ software which is freely available at (https://imagej.nih.gov/ij/download.html). The UV-vis diffuse reflectance of the LBC pellet was measured using Perkin Elmer Lambda 19 UV/Vis/NIR spectrophotometer. The optical bandgap was determined using the Kubelka Munk equation and Tauc plot. A Cameca SX-Five Field Emission Electron Probe Microanalyzer (FE-EPMA) with five 160 mm diameter Wavelength Dispersive Spectrometers (WDS) was used to measure the chemical composition of LBC and Sr-doped LBC (Detailed analysis approach provide in **SI, Section 3**). The electrical conductivity was measured in stagnant air using a four-probe DC conductivity measurement technique. Silver paint and platinum wires were used to make the electrical connection. Temperature was controlled using a Eurotherm 32H8 programmable temperature controller. Current-voltage characteristics were measured using a Keithley 6221 power supply and a 2182A nanovoltmeter. Three independent measurements on LBC and SLBC sample are performed and consistent values of conductivities and activation energies are obtained, as present in **Figure S8**. The mean value from the three measurements is presented in the main text as the experimental activation energy for LBC and SLBC.

## Supporting Information

Supporting Information is available Online.

## Author Contributions

D.M. and R.J. guided and supervised the project. J.M. performed the *ab initio* studies, analyzed the results, and wrote the manuscript. M.S.S. performed the material synthesis, XRD, SEM characterization, and conductivity measurements. L.E.S. performed the materials similarity analysis. W.O.N. performed the EPMA analysis. J. L. performed conductivity investigation and contributed to scientific discussions. M.P.P provided guidance on band alignment analysis and contributed to scientific discussions. All authors reviewed, edited, and approved the manuscript prior to submission.

## Acknowledgements

This work was funded by the US Department of Energy (DOE), Office of Science, Basic Energy Sciences (BES), under Award # DE-SC0020419. This work used the computational resources from the Advanced Cyberinfrastructure Coordination Ecosystem: Services & Support (ACCESS) program, which is supported by National Science Foundation grants #2138259, #2138286, #2138307, #2137603, and #2138296, and the computational resources provided by the Center for High Throughput Computing[65] at University of Wisconsin-Madison.

## Table of Contents

Through analyzing over 62k oxygen-containing compounds, the $MBi_2O_4X$ (M=rare-earth element, X=halogen element) structure family was identified as an ultra-fast oxygen conductor. This structure allows efficient oxygen ion diffusion through vacancies with a very low energy barrier of 0.1 eV, potentially enhancing the performance of fuel cells, solid-oxide air batteries, and other devices utilizing fast solid state oxygen transport.

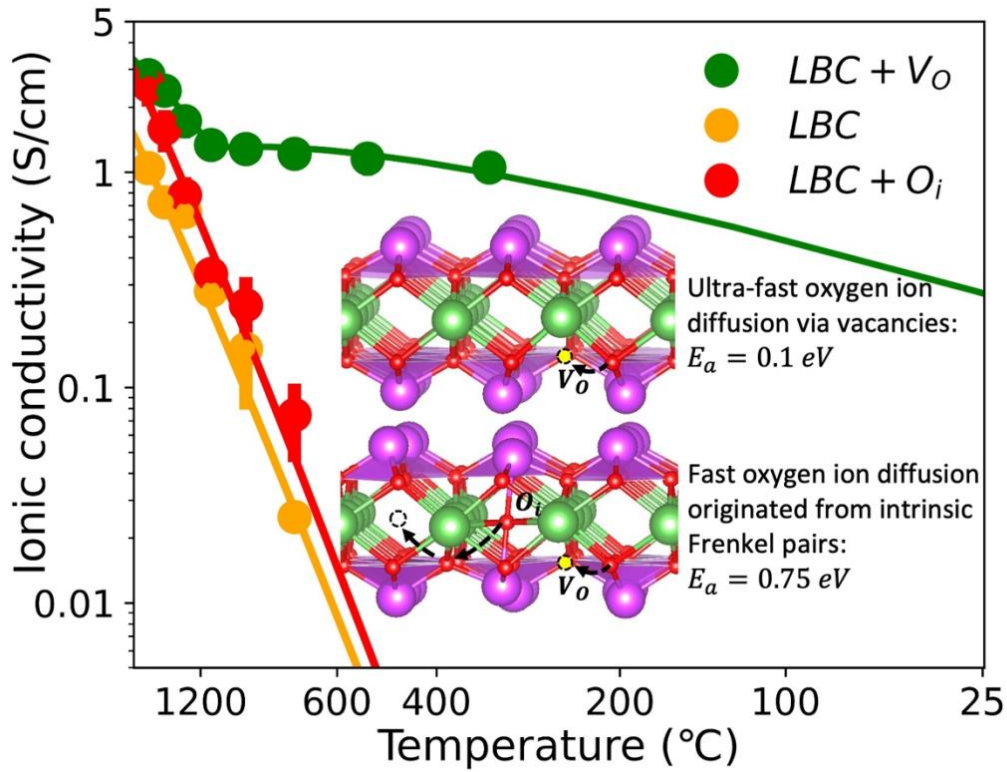

# Ultra-fast oxygen conduction in Sillén Oxychlorides

Jun Meng[1*], Md Sariful Sheikh[1], Lane E. Schultz[1], William O. Nachlas[2], Jian Liu[3], Maciej P. Polak[1], Ryan Jacobs[1], Dane Morgan[1,*]

[1]Department of Materials Science and Engineering, University of Wisconsin Madison, Madison, WI, USA.
[2]Department of Geoscience, University of Wisconsin Madison, Madison, WI, USA.
[3]National Energy Technology Laboratory, Morgantown, WV, USA.

## Table of Contents

## Section 1. Selected filters and criteria used for materials screening

Having an entry in the Inorganic Crystalline Structure Database (ICSD) database (which implies it has been made and measured) and energy above the convex hull ($E_{hull}$) were selected to assess the material's stability. Materials not present in the ICSD database or materials with an energy above the convex hull ($E_{hull}$) greater than 0 meV/atom were excluded. While this stability criterion is particularly stringent, as many materials having $E_{hull} > 0$ meV/atom can still be synthesized and used in a practical application, this criterion minimizes the risk of undesirable false positive predictions and aims to increase the likelihood of successful experimental synthesis. Additionally, we excluded materials that have similar valence electron counts to BiO₂ to enhance novelty and those with oxygen p-band center below -3.0 eV relative to the Fermi level to favor materials more prone to oxygen vacancy formation. Specifically, BiO₂ has 5 valence electrons for metal elements, and 12 valence electrons for non-metal elements. Materials with valence electron count ratio (metal/non-mental) equal to fluorite BiO₂ (5:12) were removed, again to increase novelty of the resulting candidates. In addition, the position of oxygen p-band center was used as another descriptor, which is a proxy of metal-oxygen bond strength and correlated with the formation energy of oxygen vacancies in perovskites.[1,2] In perovskites, the correlation shows that formation energy of oxygen vacancies decreases as the oxygen p-band center distance from the Fermi level increases (i.e., becomes less negative). Notably, in perovskites the formation energy of oxygen vacancies becomes positive (i.e., unfavorable energetically) when the p-band center goes below -3.0 eV relative to the Fermi level.[2] Therefore, we excluded materials with an oxygen p-band center below -3.0 eV to narrow down the materials of interest from our consideration. With applying these filters, we obtained a materials list ranking by its Euclidean distance to fluorite BiO₂, from lowest to highest, where the distance is measured from features based on x-ray diffraction and radial distribution functions (see main text). The top 10 nearest neighboring materials are listed in **Table S1**. The full list of materials is provided in Figshare.

## Section 2. *Ab initio* calculations

### 2.1 Defect formation energy

The formation energy of $V_O$ and $O_i$ are calculated by

$$E_f = E_{def} - E_{perf} \pm \mu_O + q(E_{VBM} + E_{Fermi}) + E_{corr} \quad (1),$$

where the $E_{def}$ is the total energy of the supercell with the defect, $E_{perf}$ is the total energy of the corresponding perfect supercell. $\mu_O$ is the chemical potential of oxygen and $q$ is the charge of the defect. The operator of $\mu_O$ is + for vacancy and – for interstitial, respectively. $E_{VBM}$ is the energy of the valence band maximum (VBM) of the pure material and $E_{Fermi}$ is the Fermi energy level. $E_{corr}$ is the energy correction that compensates for the electrostatic interactions arising from the periodic image of the defect. The FNV correction[3,4] was used to account for the electrostatic interaction. The formation energy of the Frenkel pair ($V_O - O_i$) is calculated as

$$E_f = E_{def} - E_{perf} \quad (2).$$

The O chemical potential was calculated by using a combination of DFT calculated total energies and experimental thermodynamic data for O₂ gas at air condition[5] with the following form[6]

$$\mu_O = \frac{1}{2}\left[E_{O_2}^{VASP} + \Delta h_{O_2}^0 + H(T, P^0) - H(T^0, P^0) - TS(T, P^0) + kT\ln\left(\frac{P}{P^0}\right) - \left(G_{O_2}^{s,vib}(T) - H_{O_2}^{s,vib}(T^0)\right)\right] (3),$$

where $E_{O_2}^{VASP}$ is the *ab initio* calculated energy of an $O_2$ gas molecule, $\Delta h_{O_2}^0$ is a numerical correction that takes into account the temperature increase of $O_2$ gas from 0 K to $T^0$, the contribution to the enthalpy at $T^0$ when oxygen is in the solid phase, and the numerical error in overbinding of the $O_2$ molecule in DFT. $\Delta h_{O_2}^0$ is obtained from comparing calculated formation energies and experimental formation enthalpies of numerous oxides, we used $\Delta h_{O_2}^0$ =0.70 eV/O from Ref.[7]. $H(T, P^0)$ and $H(T^0, P^0)$ are the gas enthalpy values at standard and general temperatures $T^0$ and $T$, respectively. In this case, $T^0$ is 298 K and $T$ refers to the studied temperature (e.g. 300 K, 873K, and 1073K). $TS(T, P^0)$ is the gas entropy, and the logarithmic term is the adjustment of the chemical potential for arbitrary pressure, where $P$ and $P^0$ are the referenced pressure and the standard pressure, respectively. In this case, the referenced pressure is 0.2 atm. The $\left(G_{O_2}^{s,vib}(T) - H_{O_2}^{s,vib}(T^0)\right)$ term accounts for the solid phase vibrations, which are approximated with an Einstein model with an Einstein temperature of 500 K.[8] Note that including this term effectively puts a solid phase contribution into the gas phase but it is convenient as it puts all the temperature dependence of Eq. (1) into $\mu_O$. Using Eq. (3), the O chemical potential at 300 K, 873K, and 1073K is determined as -4.55 eV, -5.17 eV, and -5.55 eV, respectively.

As the defect formation energies were all calculated at the GGA-PBE level, the bandgap values are underestimated. When analyzing the charged defect formation energy as a function of the Fermi energy, a recently developed modified band alignment (MBA) method[9] was used to conduct the band alignment from GGA-PBE to the HSE level, which results in a shift dependent on the bandgap of the material and on the position of a particular defect state within the gap. Hence the defect formation energy is calculated by

$$E_f^{corr} = E_f + qE_{shift} + \beta(E_f^{E_{Fermi}=0} - E_f^{q=0} - q(1-\delta)E_{gap}) + \gamma (4).$$

where the $E_f$ is the defect formation energy calculated by Eq. (1). $E_{shift}$ is the energy difference between the bandgap calculated by GGA-PBE and HSE. $E_{shift} = E_{gap}^{GGA-PBE} - E_{gap}^{HSE}$. $E_f^{E_{Fermi}=0}$ is the formation energy $E_f$ at Fermi level of 0 eV. $E_f^{q=0}$ is the defect formation energy of the non-charged defect. $E_{gap}$ is the HSE calculated bandgap. $\beta = -0.14$, $\delta = 0$, and $\gamma = 0.839$ are empirical values that were determined in the previous study.[9]

## 2.2 Diffusivity obtained from fitting Einstein relation

To enhance the statistical reliability of the diffusivity calculations from AIMD simulations, we performed a procedure fitting the Einstein relation of the MSD–$\Delta t$ curve to minimize the fitting errors. The approach involves partitioning a long-time AIMD trajectory into shorter-time ($\Delta t$) segments, then the diffusivity for each of these shorter-time ($\Delta t$) segments was calculated. This methodology was designed to exclude the ballistic region at small $\Delta t$ and the poor statistic at large $\Delta t$.[10] For each trajectory, we calculated a set of diffusivities within time intervals [$t_i$, $t_i+\Delta t$], where $t_i$ = [0 ps, 3 ps, 6 ps, 9 ps, …, ($t-\Delta t$)]. As these values are highly correlated, we kept only values that are separated by at least the correlation time, which was calculated by the autocorrelation method described in reference.[11] We denote the diffusion coefficient for one partition *i*, of one AIMD trajectory *j*, at one temperature *T*, by $D_{T,i,j}$. The mean of the diffusivities calculated on these partitions was then calculated and used as the best estimate of the diffusion coefficient for the AIMD trajectory *j*, denoted $D_{T,j} = \frac{1}{n}\sum_{i=1}^{n} D_{T,i,j}$. Here, *n* is the number of partitions. Then, we calculated the mean and standard deviation of these entirely uncorrelated $D_{T,j}$ values to obtain

$D_T = \frac{1}{m}\sum_{j=1}^{m} D_{T,j}$ (5),

$D_T^{stdev} = \sqrt{\frac{1}{m-1}\sum_{j=1}^{m}(D_{T,j} - D_T)^2}$ (6), and

$D_T^{sem} = D_T^{stdev}/\sqrt{m}$ (7),

where $m$ is the number of uncorrelated trajectories, $D_T$ is the diffusivity at temperature $T$, $D_T^{stdev}$ represents the standard deviation of the population, and $D_T^{sem}$ represents the standard error in the mean of the $D_{T,j}$, which is also called the standard deviation of the mean $D_T$. $D_T$ and $D_T^{sem}$ were utilized in the Arrhenius fitting for the migration barrier.

For the case of oxygen vacancy, we conducted 4 independent AIMD simulation trajectories with different initial seeds at temperatures ranging from 600 K to 2000 K, with a step of 200 K. For each temperature, these 4 AIMD trajectories spanned a range of time length, ranging from 207 ps to 300 ps. We have chosen $\Delta t$ of 90 ps at T=600 to 1000 K, and 120 ps at T=1200 to 2000 K, respectively. The diffusivities obtained at temperatures from 600 K to 2000 K were utilized to fit the Arrhenius relation.

For the case of intrinsic LBC, we conducted 2 independent AIMD simulation trajectories with different initial seeds from 1000 K to 2000 K in 200 K increments. These AIMD simulations spanned a range of time length from 230 ps to 390 ps We have chosen $\Delta t$ of 150 ps for each temperature. The diffusivities obtained at temperatures from 1200 K to 2000 K were utilized to fit the Arrhenius relation, as the number of hops at 1000 K was too low to yield statistically meaningful results.

For the case of LBC with oxygen interstitial, we performed 2 independent AIMD simulation trajectories with different initial seeds from 800 K to 2000 K in 200 K increments. All AIMD simulations spanned a range of time length from 230 ps to 300 ps. We have chosen $\Delta t$ of 150 ps for each temperature. The diffusivities obtained at temperatures from 1000 K to 2000 K were utilized to fit the Arrhenius relation, as the number of hops at 800 K was too low to yield statistically meaningful results.

For Sr-doped LBC, we conducted 5 independent AIMD simulation trajectories with different initial seeds from 600 K to 2000 K in 200 K increments. These AIMD simulations spanned a range of time length from 126 ps to 268 ps. We have chosen $\Delta t$ of 120 ps for each temperature. The diffusivities obtained at temperatures from 600 K to 2000 K were utilized to fit the Arrhenius relation.

## 2.3 Diffusivity predicted by a simple model of combined vacancy and interstitial transport, with contributions fit to Arrhenius equations

At high temperatures in intrinsic LBC, the concentration of diffusing defects, which arises from the formation of Frenkel pairs, is temperature-dependent and can be predicted using the Arrhenius equation

$$c_v = c_i = c_{v-i} = \exp\left(\frac{-E_f}{k_b T}\right) (8).$$

Assuming no interactions between diffusing defects, the total diffusivity can be considered as the sum of diffusions originating from oxygen vacancies and interstitials,

$$D = D_v + D_i \ (9),$$

where the vacancy diffusivity $D_v$ is determined by the hop rate of oxygen vacancy $r_v$ and its corresponding hop distance $a$,

$$D_v = r_v * a^2 = f_v * a^2 * \exp\left(\frac{-E_m^v}{k_b T}\right) (10).$$

In this context, the hop rate is defined by the attempt frequency $f_v$, estimated as $7\times10^{11}$ s$^{-1}$ (discussed below), and the migration barrier $E_m^v = 0.10 \ eV$, calculated using the CI-NEB method with GGA-PBE potential. Similarly, the oxygen interstitial diffusivity is given by

$$D_i = f_i * a^2 * \exp\left(\frac{-E_m^i}{k_b T}\right) (11),$$

where the attempt frequency $f_i$ is estimated as $5\times10^{12}$ s$^{-1}$ (discussed below), with a migration barrier $E_m^i = 0.60 \ eV$. The hop distance for both vacancy-mediated and interstitial-mediated diffusion is consistently 2.8 Å.

Initially, we fit the diffusivity data to the LBC+$V_O^{\cdot\cdot}$, at low temperatures, where the extrinsic 2.8% oxygen vacancies are the predominant defects and Frenkel pair are barely formed. We adjusted the pre-exponential factor to align with the AIMD diffusivity data and found that, with an attempt frequency for oxygen vacancy of $f_v$=$7\times10^{11}$ s$^{-1}$, the model predicted diffusivities and activation energies that are closely aligned to those observed in AIMD results. Subsequently, we applied the same attempt frequency ($f_v$=$7\times10^{11}$ s$^{-1}$) to fit the intrinsic LBC model, which includes both vacancies and interstitials from Frenkel pair formation. Using an attempt frequency for oxygen interstitials of $f_i$=$5\times10^{12}$ s$^{-1}$, the model predicts diffusivities and activation energy closely aligned with AIMD predictions. These simulations suggest that oxygen vacancies have a lower attempt frequency compared to oxygen interstitials, consistent with their lower migration barriers and the general observation that attempt frequencies and migration barriers are positively correlated.[12] In LBC, the oxygen interstitial is situated within the La layer, coordinated by four La atoms and two Bi atoms, whereas the vacancy site is surrounded by two La atoms and two Bi atoms. By differentiating the attempt frequencies for oxygen vacancies and interstitials, we are able to use the model to accurately predict diffusivities that reproduce the diffusion scenarios observed in AIMD simulations.

We also assume the Arrhneius expression for Frenkel pair formation with a unit prefactor and fit the formation energy to match the diffusivities predicted by the model and those derived from AIMD. The results demonstrate that with a Frenkel pair formation energy of 0.60 eV, this simple model accurately reproduced the diffusivities and activation energies obtained from AIMD simulations for all the three cases studied, as shown in **Figure S3**. This simple model, based on Arrhenius equations and non-interacting diffusing species, effectively explains the diffusivity values and shifts in activation energy across different extrinsic defect concentrations and temperature regimes.

## Section 3. EPMA measurements details

A Cameca SX-Five Field Emission Electron Probe Microanalyzer (FE-EPMA) with five 160 mm diameter Wavelength Dispersive Spectrometers (WDS) was used to measure the chemical composition of LBC and Sr-doped LBC (SLBC). Sintered samples of LBC and SLBC were cast in epoxy and polished to 0.25 μm with polycrystalline diamond suspension. Samples and standards were coated simultaneously with 20.0 nm Carbon. Measurements were acquired with 10 kV accelerating voltage using 20 nA beam current and fully focused beam. The following element X-ray lines were measured using the corresponding WDS diffraction crystal: O Kα on PC0, Cl Kα on LPET, Sr Lα on LTAP, La Lα on LLIF, and Bi Mα on PET. The following primary standards were used for peaking and calibration: $Bi_4Ge_3O_{12}$ (synthetic) for O and Bi, Tugtupite (natural) for Cl, $SrTiO_3$ (synthetic) for Sr, and $LaAlO_3$ (synthetic) for La. Matrix correction for quantification of O in LBC/SLBC phases involves a large absorption correction factor: 4.957 for O in LBC, 5.000 for O in SLBC. This required selection of an O primary standard with similarly high absorption correction factor (4.993 for O in $Bi_4Ge_3O_{12}$) to minimize uncertainties when converting raw X-ray counts to O concentration. The NIST FFAST table of Mass Attenuation Coefficients (MACs) was used (Chantler et al 2005).[13] Full-spectrometer scans were acquired on each standard and sample to determine unique background offsets for calibration and analysis. Each element was measured simultaneously for 60 s on peak and 30 s on each background position. An exponential background fit was applied for analysis of O on PC0. A total of 80 measurements were acquired from each sample; measurements with analytical totals outside of 98.5-101% were rejected prior to calculating average sample composition.

## Section 4. Supplementary Tables and Figures

**Table S1.** Top 10 of the nearest neighboring materials to the fluorite $BiO_2$ after applying screening filters.

| # | Materials ID | Distance (a.u.) | Formula |
|---|---|---|---|
| 1 | mp-546621 | 18.6662881 | $ErBi_2BrO_4$ |
| 2 | mp-546350 | 19.1460601 | $TmBi_2BrO_4$ |
| 3 | mp-549127 | 20.3546522 | $HoBi_2ClO_4$ |
| 4 | mp-549728 | 21.8472963 | DyZnPO |
| 5 | mp-552738 | 22.6921485 | $TmBi_2IO_4$ |
| 6 | mp-546625 | 22.6922397 | $HoBi_2BrO_4$ |
| 7 | mp-3589 | 24.1438911 | $BPO_4$ |
| 8 | mp-1087483 | 24.1444816 | ThCuPO |
| 9 | mp-6790 | 24.1687959 | $Ba_2Y(CuO_2)_4$ |
| 10 | mp-8789 | 24.2466746 | $Ca_4As_2O$ |

**Table S2.** DFT calculated bandgap (in units of eV) compared with experimental measurement.

| $E_{gap}$ | direct | Indirect |
|---|---|---|
| HSE06 | 2.38 | 2.92 |
| UV-vis | 2.65 | 3.09 |

**Table S3.** The *ab initio* study examined defect types and their formation energies in LBC (LaBi$_2$O$_4$Cl) and Sr-doped LBC (La$_{0.78}$Sr$_{0.22}$Bi$_2$O$_4$Cl) under room temperature and atmospheric $P_{O_2}$. All calculations are conducted using GGA-PBE functional. The modified band alignment (MBA) method[9] was used to shift the bandgap and defect states for charged defect. (Details described in **Section 2.1** Defect formation energy)

| | LaBi$_2$O$_4$Cl | | | La$_{0.78}$Sr$_{0.22}$Bi$_2$O$_4$Cl | |
|---|---|---|---|---|---|
| Defect type | $E_f$ (eV) at $E_{fermi}$ = 0 eV | $E_f$ (eV) at $E_{fermi}$ = 1.0 eV | $E_f$ (eV) at $E_{fermi}$ = 2.83 eV | Defect type | $E_f$ (eV) |
| $V_O$ | 2.77 | 2.77 | 2.77 | $V_{O1}$ | -0.36 |
| $V_O^{\cdot}$ | 1.48 | 2.48 | 3.88 | $V_{O2}$ | -0.28 |
| $V_O^{\cdot\cdot}$ | -0.07 | 1.93 | 4.73 | $V_{O3}$ | -0.17 |
| $O_i$ | 1.25 | 1.25 | 1.25 | | |
| $O_i'$ | 2.84 | 1.84 | 0.44 | | |
| $O_i''$ | 3.94 | 1.94 | -0.86 | | |
| $V_O - O_i$ | 0.79 | 0.79 | 0.79 | | |

**Table S4.** Comparison of the lattice vectors of DFT optimization with different functional (GGA-PBE, SCAN), and sintered LBC pellet obtained from the Rietveld refinement of XRD data and the referenced monoclinic and tetragonal phases, respectively.

| Lattice parameters | As-synthesized powder | | Sintered pellet | | DFT optimization | |
|---|---|---|---|---|---|---|
| | Monoclinic | Monoclinic Ref. [14] | Tetragonal | Tetragonal Ref. [15] | GGA-PBE | SCAN |
| a | 5.5881(9) | 5.5924(10)Å | 3.9863(3) | 3.9547 (1) | 3.99 | 3.94 |
| b | 11.5492(19) | 11.5490(2)Å | 3.9863(3) | 3.9547 (1) | 3.99 | 3.94 |
| c | 9.0044(16) | 9.0180(2)Å | 9.1063(9) | 9.1275 (3) | 9.28 | 9.09 |
| ç | 90 | 90 | 90 | 90 | 90 | 90 |
| β | 90.341(20) | 90.460 (11) | 90 | 90 | 90 | 90 |
| γ | 90 | 90 | 90 | 90 | 90 | 90 |

**Table S5.** Stoichiometry of Bi$_2$LaO$_4$Cl and Sr-doped Bi$_2$LaO$_4$Cl obtained from EPMA compared to the ideal values. The stochiometric values are shown as 1σ ranges from the sample mean.

| | La | Sr | Bi | O | Cl |
|---|---|---|---|---|---|
| Ideal LBC | 1 | - | 2 | 4 | 1 |
| synthesized LBC | [0.97, 1] | - | [1.97, 2.03] | [3.84, 3.93] | [1.01, 1.04] |
| Ideal Sr-doped LBC | 0.95 | 0.05 | 2 | 3.975 | 1 |
| synthesized Sr-doped LBC | [0.90, 0.95] | [0.01, 0.01] | [1.95, 2.05] | [3.76, 3.93] | [0.99, 1.04] |

**Table S6.** Lattice vectors of as-synthesized Sr-doped LBC powder and sintered pellet obtained from the Rietveld refinement of XRD data.

| Lattice parameters | As synthesized powder | Sintered pellet |
|---|---|---|
| | Monoclinic | Tetragonal |
| a | 5.5935(5) Å | 3.9923(9) Å |
| b | 11.5515(10) Å | 3.9923(9) Å |
| c | 9.0192(9) Å | 9.1553(3) Å |
| $\alpha$ | 90 | 90 |
| $\beta$ | 90.368(10) | 90 |
| $\gamma$ | 90 | 90 |

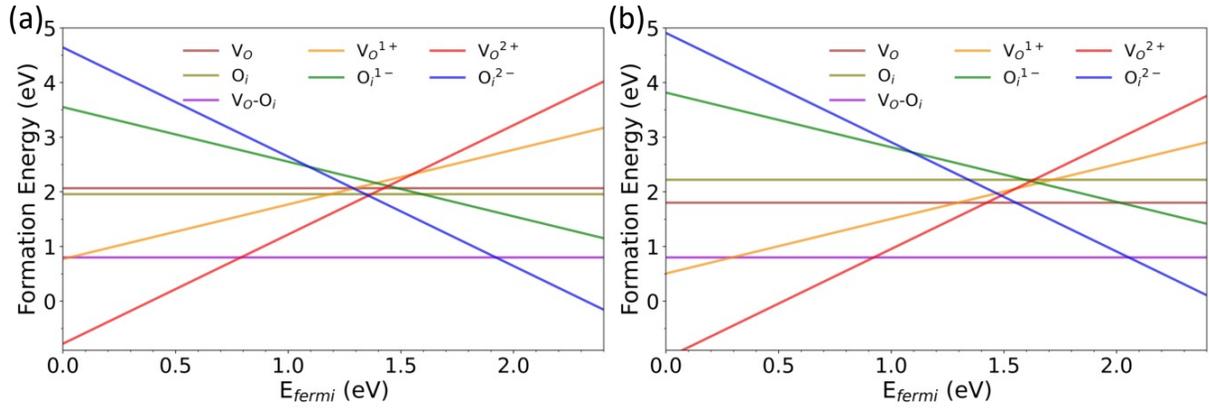

**Figure S1.** Defect formation energies as a function of Fermi level for intrinsic defects considered in LBC under atmospheric $P_{O_2}$ at (a) 873K and (b) 1073K, respectively. The temperature dependence in these levels comes from the temperature dependence of the $\mu_O$ term in Eq. (3). Full computational details are provided in **Methods 2** in the main manuscript.

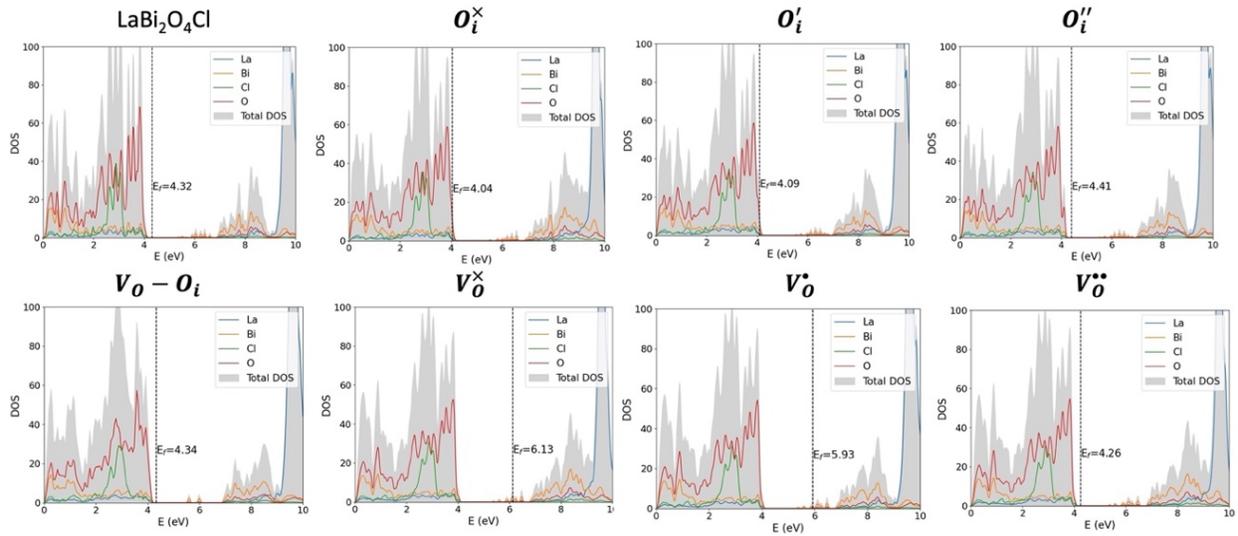

**Figure S2.** The density of states calculated with GGA-PBE functional of pure LaBi$_2$O$_4$Cl system and LaBi$_2$O$_4$Cl with the presence of oxygen interstitial ($O_i$), Frenkel pair ($V_O - O_i$), or oxygen vacancy ($V_O$).

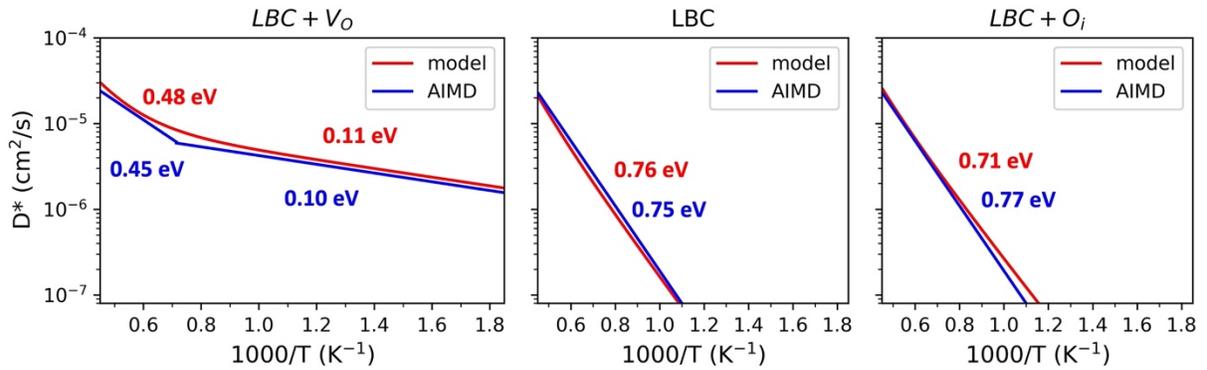

**Figure S3.** Oxygen self-diffusivity of LBC with 2.8% $V_O$, intrinsic LBC, and LBC with 2.8% $O_i$ predicted by the simple defect concentration model compared to AIMD results.

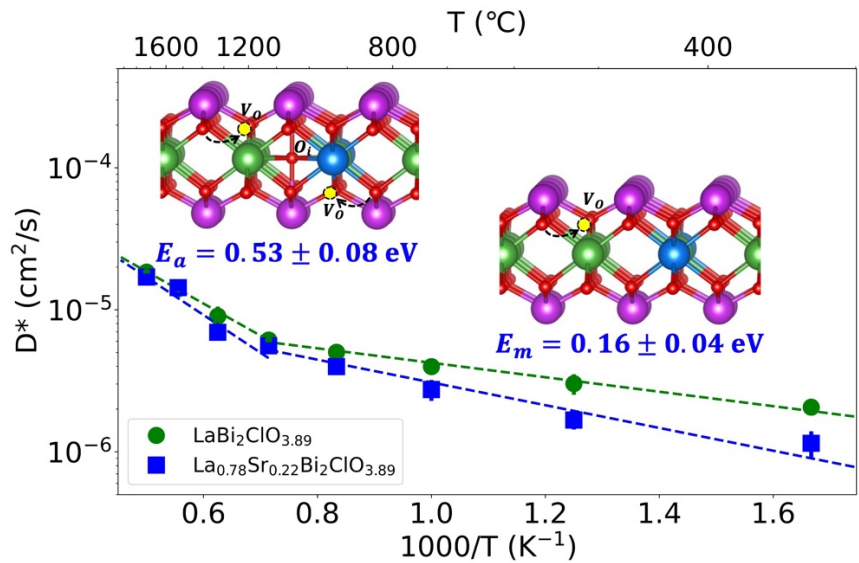

**Figure S4.** Arrhenius plot of oxygen self-diffusivity $D^*$ in Sr-doped LBC with 2.8% oxygen vacancies calculated by AIMD simulation using GGA-PBE functional, compared to the LBC with 2.8% extrinsic oxygen vacancies system. The inserted images illustrate the diffusion mechanism at high and low temperature regimes in Sr-doped LBC.

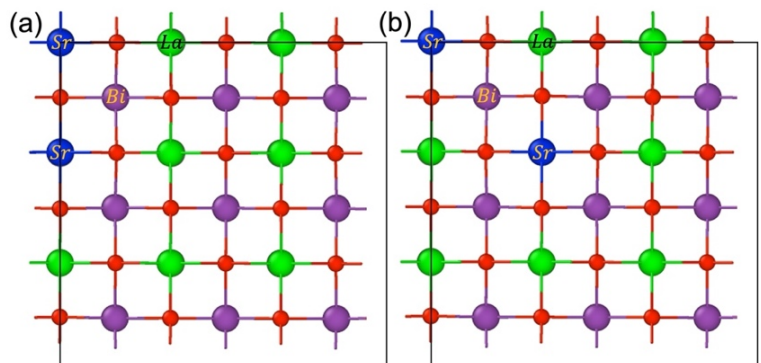

**Figure S5.** The two considered configurations of the Sr dopants in LBC for DFT calculations. (a) Configuration 1 and (b) Configuration 2.

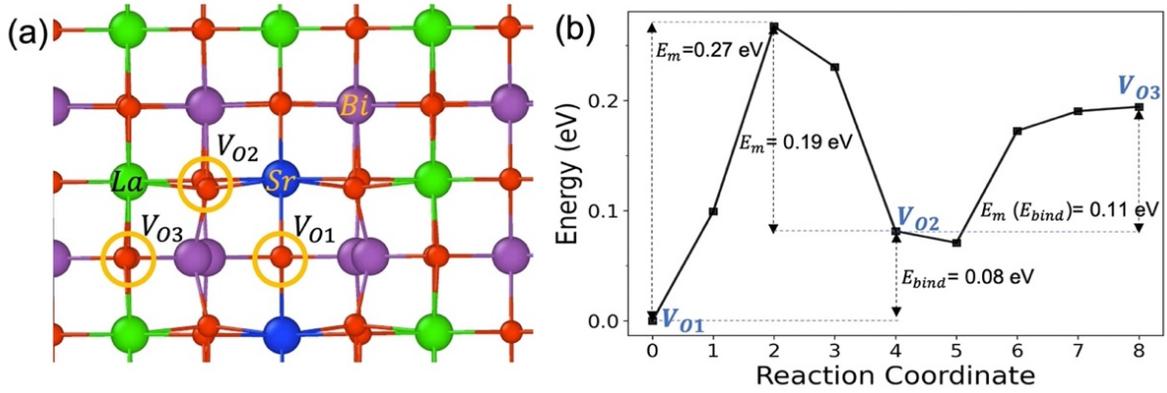

**Figure S6.** CI-NEB studies of migration energy barriers for all possible diffusion pathways in La$_{0.78}$Sr$_{0.22}$Bi$_2$O$_{3.89}$Cl, calculated using GGA-PBE functional. The energy differences of 0.08, 0.11, and 0.19 eV between different vacancy sites represent the binding energies between oxygen vacancies and Sr dopants.

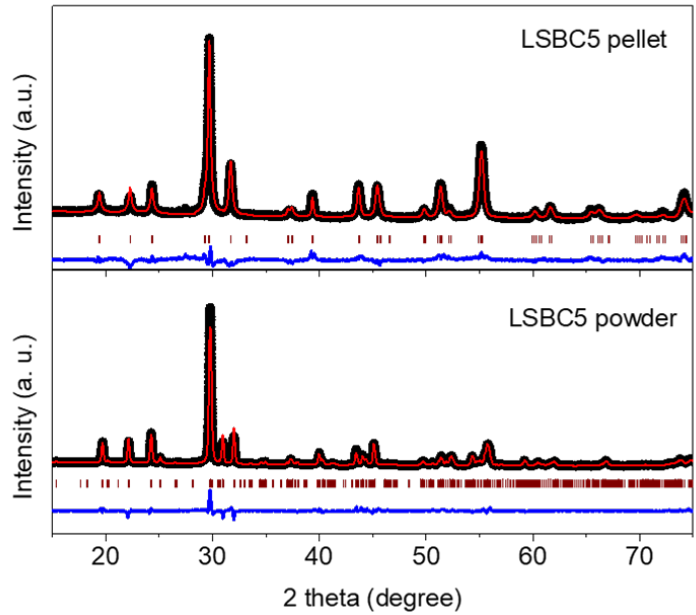

**Figure S7.** X-ray powder diffraction data of the synthesized Sr-doped LBC power fit to the monoclinic phase, and the sintered Sr-doped LBC pellet fit to the tetragonal phase.

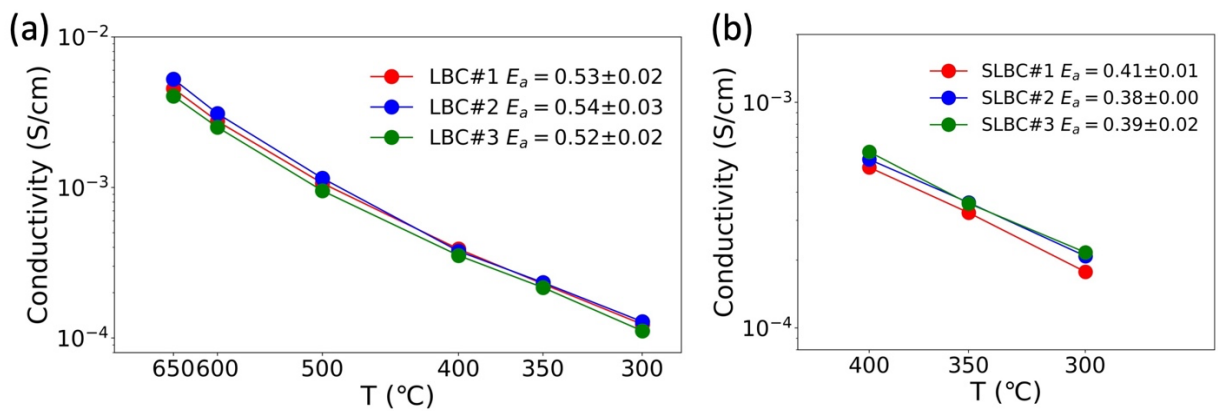

**Figure S8.** Conductivity of (a) LBC and (b) SLBC measured by 4-probe DC method from three independent measurements.